\documentclass[aps,prb,longbibliography,amsmath,amssymb,superscriptaddress,twocolumn,10pt]{revtex4-2}
\usepackage{standalone}
\usepackage{txfonts}% Physical Review B
\usepackage{graphicx}% Include figure files
\usepackage{dcolumn}% Align table columns on decimal point
\usepackage{bm}% bold math
\usepackage{array}
\usepackage[dvipsnames]{xcolor}
\usepackage[%  
    colorlinks=true,
    pdfborder={0 0 0},
    linkcolor=blue
]{hyperref}
\usepackage{chngpage}
\usepackage{appendix}
\usepackage{xspace}
\usepackage{comment}

\graphicspath{{Figures/}}

\begin{document}

\def\avs {AV$_3$Sb$_5$\xspace}
\def\rvs {RbV$_3$Sb$_5$\xspace}
\def\cvs {CsV$_3$Sb$_5$\xspace}
\def\avs {$A$V$_3$Sb$_5$\xspace}
\def\cvss {CsV$_3$Sb$_{5-x}$Sn$_x$\xspace}
\def\kvs {KV$_3$Sb$_5$\xspace}
\def\V {$^{51}$V\xspace}
\def\Sb {$^{121}$Sb\xspace}

\title {Competing lattice structures induced by Sn substitution in \cvs}

\author{\hspace{1mm}Anshu Kataria}
\email[]{kataria96anshu@gmail.com}
\affiliation{Dipartimento di Scienze Matematiche, Fisiche e Informatiche, Universit\`a di Parma, I-43124 Parma, Italy}

\author{\hspace{1mm}Ilija K.~Nikolov}
\affiliation{Department of Physics, Brown University, Providence, Rhode Island 02912, USA}

\author{\hspace{1mm}Armando Consiglio}
 \affiliation{Dipartimento di Fisica e Astronomia ``A. Righi'', Universit\`a di Bologna, I-40127 Bologna, Italy }
\affiliation{Istituto Officina dei Materiali, Consiglio Nazionale delle Ricerche, Trieste I-34149, Italy}

\author{\hspace{1mm}Giuseppe Allodi}
\affiliation{Dipartimento di Scienze Matematiche, Fisiche e Informatiche, Universit\`a di Parma, I-43124 Parma, Italy}

\author{\hspace{1mm}Ginevra Corsale}
\affiliation{Department of Physics, Brown University, Providence, Rhode Island 02912, USA}
\affiliation{Dipartimento di Fisica e Astronomia ``A. Righi'', Universit\`a di Bologna, I-40127 Bologna, Italy }

\author{\hspace{1mm}Andrea Capa Salinas} 
\affiliation{
Materials Department, University of California Santa Barbara,
Santa Barbara, California 93106, USA}

\author{\hspace{1mm}Stephen D. Wilson}
\affiliation{
Materials Department, University of California Santa Barbara,
Santa Barbara, California 93106, USA}

\author{\hspace{1mm}Domenico Di Sante}
 \affiliation{Dipartimento di Fisica e Astronomia ``A. Righi'', Universit\`a di Bologna, I-40127 Bologna, Italy }

\author{\hspace{1mm}Vesna F. Mitrovi{\'c}}
\affiliation{Department of Physics, Brown University, Providence, Rhode Island 02912, USA}
\affiliation{Brown Center for Theoretical Physics and Innovation, BCTPI, Brown University, Providence, RI 02912-1843, USA}

\author{\hspace{1mm}Samuele Sanna}
 \affiliation{Dipartimento di Fisica e Astronomia ``A. Righi'', Universit\`a di Bologna, I-40127 Bologna, Italy }

\author{\hspace{1mm}Pietro Bonf\`a}
\email[]{pietro.bonfa@unimore.it}
\affiliation{Dipartimento di Fisica, Informatica e Matematica, Universit\`a di Modena e Reggio Emilia, Via Campi 213/a, 41125 Modena, Italy}
\affiliation{CNR-NANO S3—Istituto Nanoscienze, I-41125 Modena, Italy }

\date{\today}

\begin{abstract}

Understanding the effect of chemical substitution on competing phases of kagome metals is crucial for disentangling the interplay between local structural distortions and electronic instabilities.  In \cvs, Sn substitution strongly modifies the electronic phase diagram, yet the microscopic mechanism driving this remains unclear.
%Vanadium-based kagome metals exhibit intertwined electronic states, including charge density wave, nematicity and superconductivity. Chemical substitution and external pressure can be used to further tune the electronic states, leading to suppression of charge order and appearance of double-dome superconductivity. These electronic states are extremely sensitive to weak external stimuli, including strain and small magnetic fields, preventing a clear distinction between intrinsic electronic behavior and perturbation-induced effects.
%In this context, to characterize the role of Sn chemical substitution in \cvs, 
%in the vanadium-based kagome metals, 
Here, we combine $^{121}$Sb nuclear quadrupole resonance (NQR) measurements and density functional theory calculations to investigate the atomic-scale effects of Sn substitution in CsV$_3$Sb$_{5-x}$Sn$_x$.
At low Sn concentrations, the observed satellite NQR peaks exhibit signatures of local structural distortion induced by Sn substitution, qualitatively consistent with our computational analysis. These impurity-induced features persist across the entire experimentally investigated doping range, up to $x$ = 0.65, and remain observable up to room temperature.
%This distortion is far from trivial and extends to multiple nearest neighbors of the dopant.
For $x=1$, the fully doped idealized case, the estimated dynamical instabilities of the kagome lattice suggest the stabilization of two nearly energy-degenerate equilibrium structures characterized by V-trimers, distinguished by a zero- or $\pi$-phase shift between adjacent layers along the $c$-axis.  Together, these results show that Sn substitution drives a complex interplay between local impurity-induced distortions and competing structural instabilities in vanadium-based kagome compounds.
\end{abstract}

\maketitle
\section{Introduction}

Vanadium-based kagome metals \avs ($A$ = K, Rb, Cs)~\cite{CVS_Z2,luo2022electronic,kagome.first} provide a new platform to investigate correlated electronic phases and intertwined orders, including charge density wave (CDW), superconductivity (SC), nematic order, and potential time-reversal symmetry broken states\mbox{~\cite{mielke2022time,nie2022charge,li2023unidirectional,trsbkagome,PhysRevResearch.4.023244,Zhao_2021,1g9n-wm38,wang2023quantum,neupert2022charge}}. The CDW state in these metals is primarily associated with Fermi surface instabilities \cite{jiang2021unconventional,PhysRevLett.127.046401,optical,kang2022twofold,PhysRevLett.128.036402,wilson2024v3sb5}, although recent investigations have also proposed the possible role of the electron-phonon interaction \cite{PhysRevLett.127.217601,dftcdw,xraycdw,PhysRevB.105.L140501,enzner2025phonon}. 
%Notably, first-order, order-disorder type transitions emerging from configurational entropy have also been proposed at $T_{CDW}$ \cite{subires2023order,yao2024nature,PhysRevMaterials.8.064002}. 
Structurally, the CDW phase involves lattice distortions mostly localized within the vanadium kagome plane, stabilizing a Star of David~(SoD) or Tri-Hexagonal~(TrH) pattern characterized by the 3q-breathing mode, with 0 or $\pi$ phase shift between neighboring layers. These 3D CDW orders are linked to unstable phonon modes at the $M$ and $L$ high-symmetry points in the momentum space\mbox{~\cite{PhysRevB.104.214513,PhysRevLett.127.046401,PhysRevB.107.205131,PhysRevB.105.235134}}. %highlighting the strong interplay of lattice and electronic instabilities in these systems.  

External tuning parameters, such as pressure and chemical doping, can be used as an effective tool to understand the nature of competing phases or electronic states in these materials, as they can modify the Fermi surface and local structural environment~\cite{PhysRevB.105.165146}. In \cvs, both chemical substitution and applied pressure suppress the long-range charge order and induce a double-dome SC feature~\cite{Chen2021,Wang2022,Feng2023,Zheng2022,CVSSn,Kautzsch2023, kang2023charge,SciPostPhys.12.2.049}.  Further, Sn substitution at the Sb site has been reported to produce a quasi-one-dimensional incommensurate stripe-like modulation and to suppress long-range charge order beyond a critical doping level ~\cite{Kautzsch2023, Huai2025}.
In contrast, recent investigations indicated the presence of short-ranged CDW correlations well beyond this critical doping level \cite{ilija_manuscript,deng2026observation,Kongruengkit2026}, and are found to persist up to room-temperature (RT)~{\cite{ilija_manuscript,deng2026observation}}.
Additionally, ultrafast reflectivity studies indicated Sn doping-induced modifications of the CDW structure, suggesting a transition from the pristine SoD+TrH mixture to the $\pi$-shifted TrH phase~\cite{deng2025coherent}. 
Taken together, these results present a complex interplay between chemical disorder and electronic instabilities.
%not only near the CDW transition but also in regime where long-range order is absent, $i.e.$, at room temperature.
%At the same time, a room-temperature powder X-ray studies showed the doped system retains the pristine kagome structure \cite{CVSSn}. 
 
Finally, the CDW state in pristine \cvs is known to be extremely sensitive to mechanical strain and applied magnetic fields~\cite{Guo2024, Frachet2024}, indicating that seemingly small perturbations can act as a crucial tuning parameter between multiple nearly degenerate electronic states. 
%suggesting that experimental response results from a competition between nearly degenerate states.
%the near-degeneracy of multiple competing states with delicate energy balance
%indicating a finely balanced competition among the nearly degenerate structural states. 
Chemical substitution is also expected to significantly alter this balance and thereby affect the resulting ground states, thus motivating the present investigation on the hole-doped \cvs system.

In this context, we perform $^{121}$Sb nuclear quadrupole resonance (NQR) measurements on high-quality \cvss powders and density-functional theory (DFT) simulations. NQR is an effective technique to probe the local charge distribution around the nuclei due to its ability to directly measure the electric field gradient (EFG) at a nuclear site~\cite{abragam1961principles,das1958nuclear}.
This enables us to investigate localized charge distributions and rearrangements around the substituted atom that are not directly accessible to bulk diffraction measurements. 
Our $^{121}$Sb NQR measurements on \cvss reveal an in-plane impurity-induced local distortion, manifested as satellite peaks around the main peak in the NQR spectra. These distortions are substantial and, for full substitution ($x = 1$), give rise to a new dynamical instability of the lattice, resulting in two nearly degenerate lattice structures for CsV$_3$Sb$_4$Sn.
 
\section{Results}
\subsection{Room-temperature $^{121}$Sb NQR spectra}
\begin{figure}
\includegraphics[width=0.45\textwidth]{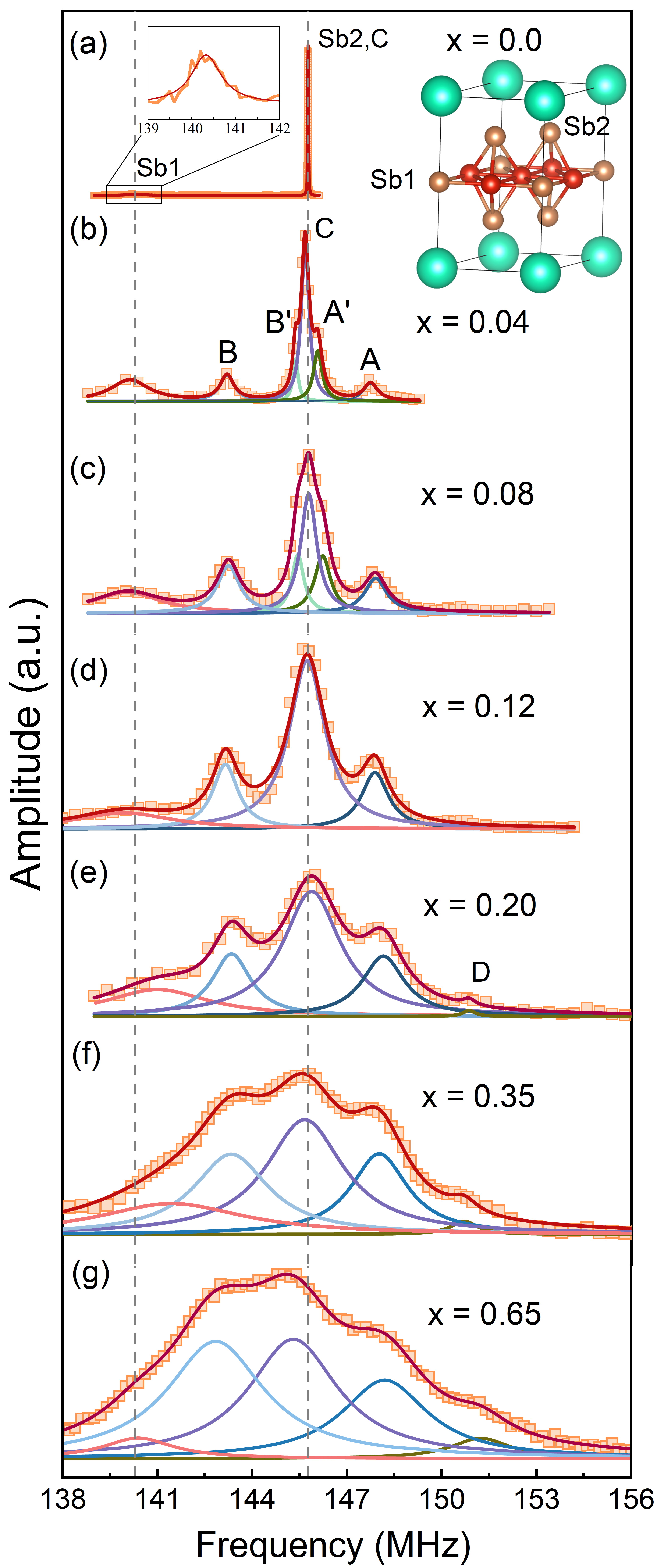}
\caption{Room temperature $^{121}$Sb NQR spectra of powdered \cvs sample as a function of Sn doping for the ${5/2 \leftrightarrow 3/2}$ transition. Inset shows the RT hexagonal structure of the pristine system and an enlarged view of the broad Sb1 peak. The satellite resonance peak frequencies $\nu_Q$ of Sb2 remain unaffected with increasing doping concentration and remain symmetric around the central C peak. Vertical dashed lines correspond to the Sb resonance frequencies for the pristine sample. The spectral profile is fitted with a multiple Lorentzian distribution, where the differently coloured fitted components represent distinct Sb environments and the solid red curve represents their cumulative fit. }
\label{fig1:RT_sb2}
\end{figure}
The RT hexagonal crystal structure of \cvs consists of two Sb sites: Sb1 atoms lie in the kagome plane surrounded by corner-sharing V-triangles, and Sb2 atoms above and below the kagome plane forming a honeycomb lattice, as shown in the inset of Fig.~\ref{fig1:RT_sb2} \cite{kagome.first}. The two sites are in a ratio Sb1: Sb2 = 1:4.
Fig.~\ref{fig1:RT_sb2} depicts the RT $^{121}$Sb NQR spectra as a function of Sn doping in \cvss. The signal acquired for the pristine system is shown in Fig.~\ref{fig1:RT_sb2}(a). The observed two peaks at 140.1 and 145.8 MHz are attributed to the Sb1 and Sb2 nuclei, respectively, based on their relative intensities, which match the expected atomic site multiplicities. 
It is worth noting that the Sb1 peak is broader than Sb2 (inset of Fig.~\ref{fig1:RT_sb2}), suggesting a much more disordered local environment, as already reported in previous investigations on single crystals~\cite{feng2023commensurate,Mu_2022,ilija_manuscript}.
%suggesting a much more disordered local environment. This result is not sample-specific and has also 
% This inherent effect has already been reported in previous investigations employing single crystals~\cite{feng2023commensurate,Mu_2022, ilija_manuscript}.

Remarkably, with Sn doping (Fig.~\ref{fig1:RT_sb2}(b-g)), multiple pairs of satellite peaks appear in the spectra. These satellite peaks are roughly equally intense and symmetric around the central Sb2 peak (labeled C). Given their large spectral weight, we tentatively attribute them to Sb2 nuclei subjected to a different EFG, $i.e.,$ a different charge environment. This is discussed in a later section. 

Fig.~\ref{fig1:RT_sb2}(b) shows the NQR spectra for a lightly hole-doped system, $x = 0.04$. Four satellite peaks showing larger (smaller) shifts relative to the C peak can be distinguished and are labeled as A and B (A\textquotesingle~and B\textquotesingle) \footnote{A\textquotesingle ~and B\textquotesingle ~are at +0.4~MHz and -0.3~MHz, while A and B are at +2.5~MHz and -2.1 MHz, respectively, from the C peak.}. As the doping level increases to $x=0.12$ (Fig.~\ref{fig1:RT_sb2}(d)), the C peak broadens significantly due to enhanced configurational disorder. Consequently, only the satellite peaks with larger shifts (A and B) are resolved at the higher doping levels.  
Surprisingly, these satellite peaks persist even up to the highest doping concentration of $x$ = 0.65 (Fig.~\ref {fig1:RT_sb2}(e-g)). A fifth peak (labeled D) at around 151~MHz develops for $x>0.20$; however, its intensity remains very weak.

The experimental NQR spectra are fit with a sum of Lorentzian functions, with each component shown in a different color.  The cumulative fit (red) matches the experimental data (orange squares) very well (Fig.~\ref {fig1:RT_sb2}).
The dashed vertical lines represent the Sb NQR frequency for the pristine sample. 
The resonance frequency of each NQR peak and their dependence on doping concentration are shown in the Supplemental Material \footnote{See Supplemental Material at~\cite{} for the details on experimental and computation details, experimental Sb NQR frequencies as a function of doping, NQR frequencies for Sn-doping using density functional theory and phonon analysis of CsV$_3$Sb$_4$Sn in trimerized structure
%, which includes Refs.\cite{Perdew1996,rvv10,gipaw,Monkhorst1976,kagome.first,PhysRevMaterials.6.015001,DalCorso_2014,phonopy,gbrv,CVSSn}
}.

\begin{figure}
\includegraphics[width=0.43\textwidth]{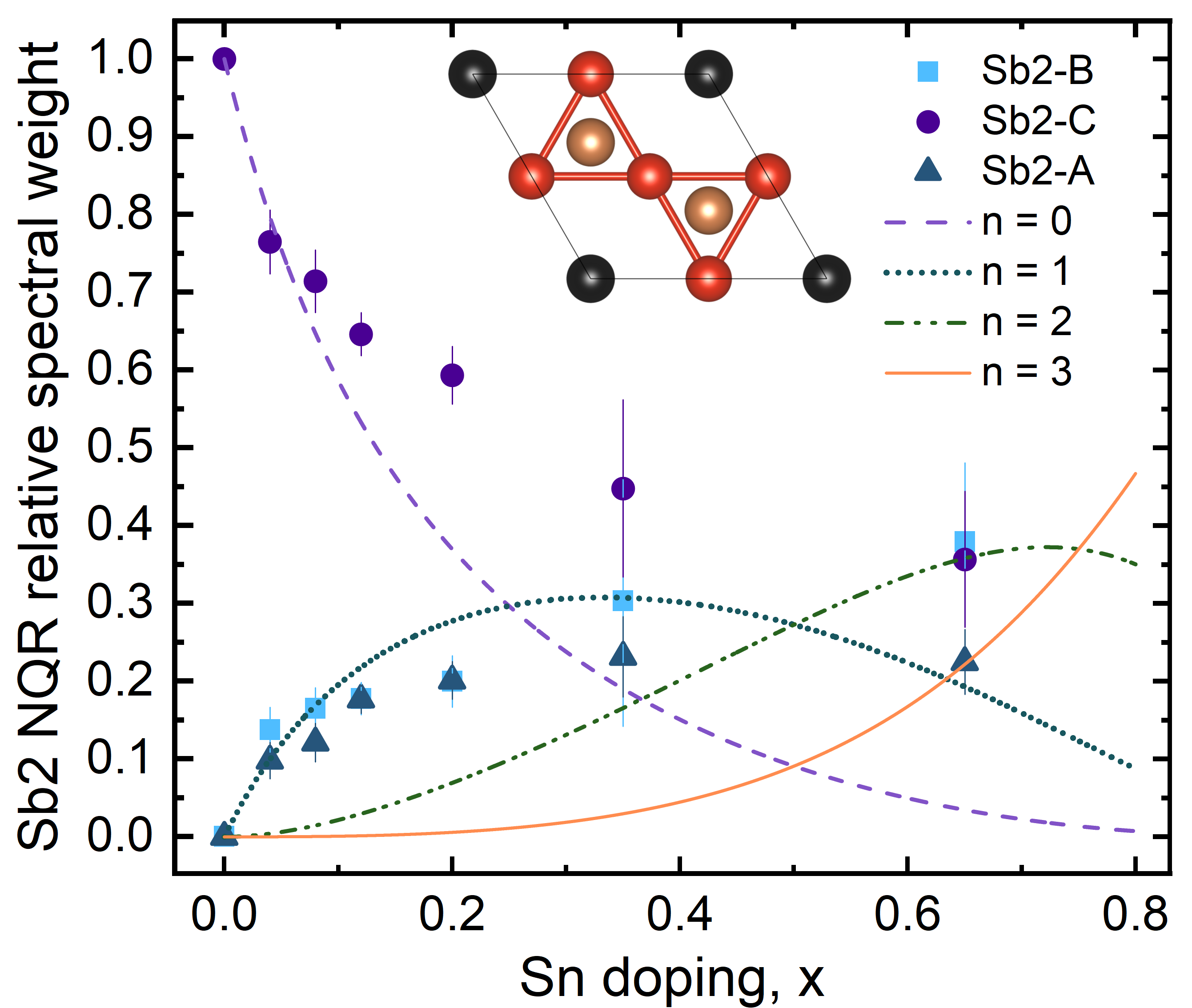}
\caption{Relative spectral weight variation of different Sb2 NQR peaks as a function of Sn doping for RT measurements. Various colored lines represent the simple binomial distribution function for Sn at the in-plane Sb1 site. The inset shows the hexagonal structure in the $ac$-plane with Sn atoms at the Sb1 site, presented by black spheres (Sn-NN). $n$ represents the number of Sn-NN in the vicinity of Sb2 atom, with $n=0-3$ corresponding to different local environments around Sb2. Experimental data can only distinguish the Sb2 atom having at most one Sn atom among the neighbors.}
\label{fig2:reltaive weight}
\end{figure}

Fig.~\ref{fig2:reltaive weight} reports the evolution of the relative spectral weight of the NQR peaks as a function of Sn doping. 
The relative weight is defined as the ratio between the spectral weight of each component with respect to the total spectral weight of all Sb2 nuclei ($i.e.,$ excluding the peak at the lowest frequency attributed to Sb1). 
The results reveal two notable features: (i) the spectral weight ratio of the satellite peaks A and B remains close to 1:1 for the entire doping series and (ii) the relative spectral weight of the central C peak decreases with doping, whereas those corresponding to the population of A and B nuclei increase.

These trends can be modeled with a simple binomial distribution accounting for the random site occupancy of Sn at Sb1 sites. In agreement with previous literature, we assume that Sn substitutes only at the in-plane Sb1 site in the kagome plane (See ref.~\cite{CVSSn} and SM).
The rationale behind this procedure is that the satellite peaks appearing in the NQR spectra could be associated with the perturbation induced by Sn substitution at Sb sites and should therefore scale with the doping concentration.
Under these assumptions, Sb2 atoms can have up to three Sn atoms among their nearest neighbors (Sn-NN) and we assume that each of these configurations produces a different signature in the NQR spectra reflecting different EFGs. The local environment of the Sb2 nuclei in the $ac$-plane is shown in the inset of Fig.~\ref{fig2:reltaive weight}.

As already mentioned, the experimental data mainly distinguish the contribution from the central C peak, and the A and B satellites. Additional contributions corresponding to other local environments may be present in the experimental data but cannot be resolved. 
The spectral-weight trends of the C, A and B peaks are therefore compared with the prediction from the binomial distribution in Fig.~\ref{fig2:reltaive weight}, where the C peak can be associated with the probability of having no Sn nearest neighbor around, whereas the A and B peaks are related to configurations with one Sn nearest neighbor. The remaining configurations are here neglected.

The experimental trends are well described by this simple model at low doping levels, supporting the conclusion that Sn randomly occupies the Sb1 site and gives rise to the NQR signature represented by the A and B peaks. 
However, with increasing $x$, 
%the spectra exhibit broadening, which might include the 
the line broadening makes it challenging to accurately determine the spectral weight of each peak, which results in deviations from the ideal binomial distribution.
In addition, at higher doping levels, the possibility of Sn occupying the Sb2 site cannot be excluded.

%\subsection {Computational results}

\subsection{Microscopic origin of NQR spectra}

In the previous section, we suggested that the satellite peaks observed in NQR spectra of the doped system can be associated with a Sn-concentration–dependent modification of the Sb2 charge environment. 
In order to address the origin of these peaks, we compute the local charge redistribution and hence the modification of NQR spectra due to the Sn substitution from first-principles simulations.

We first consider the $T=$~0~K ground-state structure, which is directly obtained by standard DFT calculations. Earlier investigations have shown that the low-temperature ground state of the pristine system is characterized by the TrH lattice distortion. This configuration is slightly lower in energy (-1.7 meV/atom) than the SoD pattern (-0.5 meV/atom) with respect to the high-temperature hexagonal lattice~\cite{PhysRevLett.127.046401}. Moreover, modulation along the $c$-axis further reduces the total energy by $\simeq$10 meV/atom~\cite{PhysRevLett.127.046401}, and depending on the alkali atom, different stacking orders are found to be nearly degenerate ~\cite{4r8x-j3nd}.
Previous studies on doped systems,  CsV$_3$Sb$_{5-x}$Sn$_x$ at $x=1$~\cite{CVSSn}, showed that the Sn preferentially substitutes at the in-plane Sb1 site. Our calculations independently confirm this site preference (See SM).
However, modeling intermediate doping levels ($0<x<1$) would require accounting for the complex energy landscape generated by both compositional disorder and the multiple nearly degenerate electronic states characterizing the pristine system (more details are also provided in the SM). For this reason, in the following we focus only on the low-doping regime of the experimental results and neglect the order or correlations along the $c$-axis, which have also been found to decrease with increasing doping concentration \cite{CVSSn}.

\begin{figure}
\includegraphics[width=0.416\textwidth]{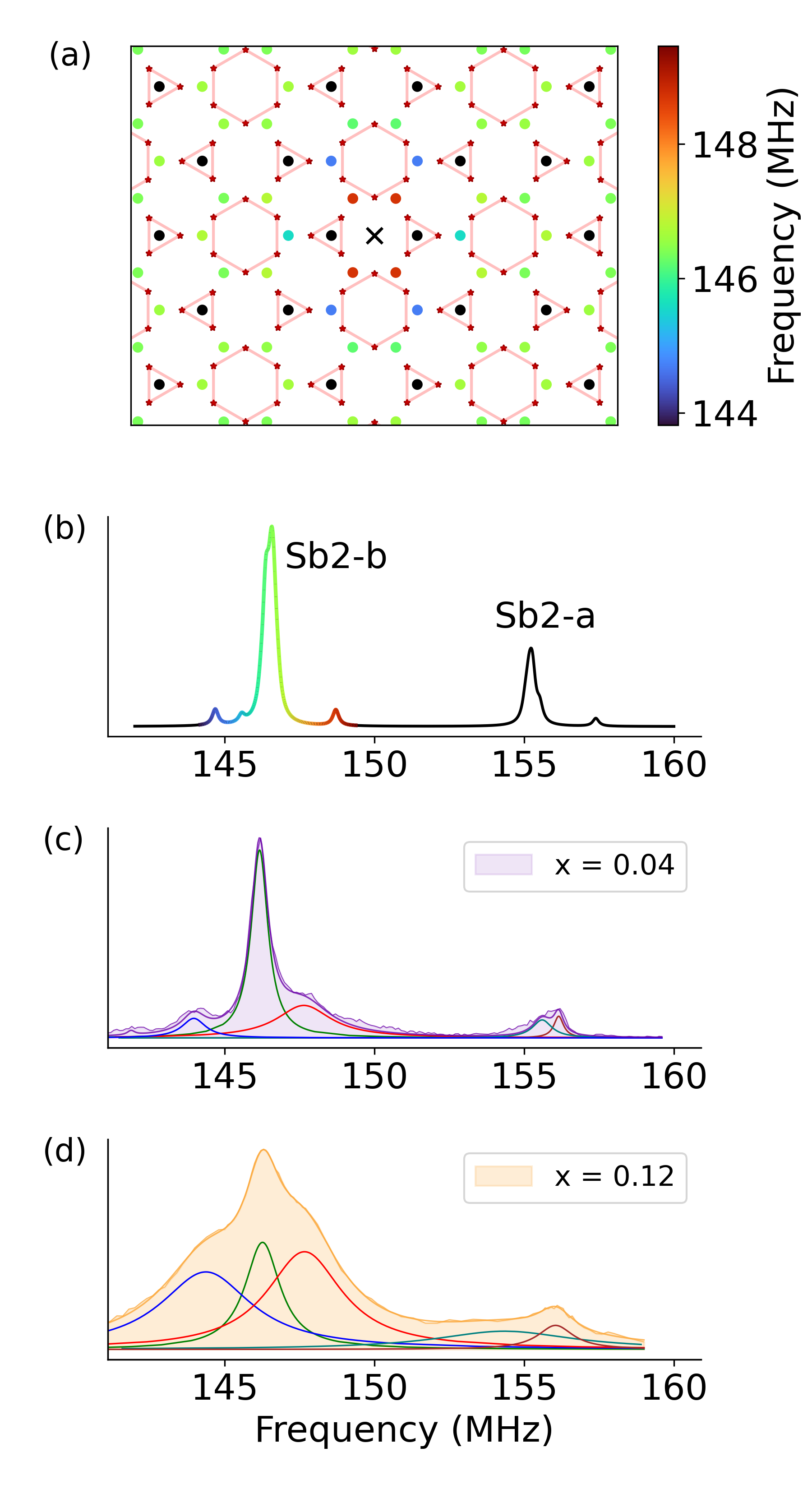}\caption{
Low-temperature $^{121}$Sb NQR spectra and computational predictions of the resonant frequencies for ${5/2\leftrightarrow 3/2}$ transition. DFT-estimated NQR frequencies of Sb2 nuclei mapped on the structural arrangement in panel (a). To better highlight the spatial distribution of the perturbation, only the Sb2-b ($i.e.,$ those in the $12o$ Wyckoff site) are considered, while the remaining Sb2-a sites are indicated as black points for clarity and displayed in an enlarged view. The V atoms are shown as dark red stars, and the Sn atom is shown by the black cross. The predicted spectra for all Sb2 atoms are shown in panel (b). Notice that all values have been shifted by 13.1 MHz to account for systematic errors in the DFT prediction. Panels (c-d) present the experimental NQR spectra of \cvss for $x=0.04, 0.12$ at T = 5 K. The experimental spectra in the shaded area are fitted with the sum of Lorentzian functions, with each component depicted by a different solid color. Notably, the color assignment of the Sb2-b peak in panels (b-d) follows the color of the corresponding Sb2-b nuclei shown in panel (a).}
\label{fig:impurity}
\end{figure}

To support the assignment of the Sn-induced NQR peaks based on the binomial analysis discussed in the previous section, from first principles, we computed the EFG at the Sb2 sites for  $x \approx 0.015$ ($8\times8\times1$ supercell), considering the TrH structure and substituting the Sn impurity at the Sb1 site outside the two V hexagons ($i.e.,$ $3f$ Wyckoff site), indicated by the black cross in Fig.~\ref{fig:impurity}(a).

In the TrH structure of the pristine material ($i.e,$ for $x=0$), the Sb2 atoms split into two inequivalent Wyckoff sites, $4h$ and $12o$, labeled as Sb2-a and Sb2-b, respectively, leading to two expected NQR peaks, according to previous results~\cite{PhysRevB.107.184106, PhysRevResearch.5.L012017}.
%These have indeed been observed previously, along with a small perturbation effect introduced by the modulation along the $c$ direction \cite{PhysRevB.107.184106}.
Experimentally, these two peaks are characteristic of the TrH configuration and are reported to persist in the whole Sn-doped series at low temperature~\cite {ilija_manuscript}. In Fig.~\ref{fig:impurity}(c-d), the experimental NQR spectra (shaded area) obtained at T = 5~K for $x=0.04$ and $0.12$ doped systems are shown with the multiple Lorentzian fit represented by solid curves. 

The predicted NQR resonance frequencies for the \mbox{5/2 $\leftrightarrow $ 3/2} transition of the  $x\approx 0.015$ case are presented in Fig.~\ref{fig:impurity} (a-b) \footnote{As also described in the caption of the figure, a constant shift of 13.1 MHz is added to account for systematic errors in the prediction based on plane wave DFT.}. 
The perturbation induced by the Sn impurity (shown as a black cross) is represented by resonance mapping onto the structural arrangement, as shown in the upper panel of Fig.~\ref{fig:impurity}(a). To better highlight this effect, we focus only on the Sb2-b~nuclei, leaving the Sb2-a~sites off scale, however showing them as black points for clarity.
As shown in Fig.~\ref{fig:impurity}(a), far from the Sn atom, the effect of the perturbation vanishes, and the Sb2-b site retains its pristine resonance frequency (green points). In contrast, Sb2-b atoms close to the Sn impurity exhibit increased and decreased frequencies relative to the Sb2-b resonance (red and blue/light blue points, respectively).
In Fig.~\ref{fig:impurity}(b), the intense peaks at $\approx$148 and 156 MHz correspond to the Sb2-b and Sb2-a~nuclei, respectively, and originate from the unperturbed Sb atoms far from Sn (green and black).  
Additional satellite peaks emerge above (red) and below (blue/light blue) the Sb2-b~major peak, corresponding to the nearest- and next-nearest-neighbor Sb2 atoms relative to the Sn dopant. The Sn impurity also affects the Sb2-a~peak, leading to an additional peak at higher frequency ($\simeq$158 MHz). Although the agreement between the computational and experimental results is only qualitative, the emergence of satellite peaks above and below the most intense Sb2 peaks provides direct evidence of the local charge modification induced by Sn substitution. This finding also supports our initial assignment of the high-temperature NQR peaks. %based on the binomial distribution analysis.
These satellite peaks survive up to room temperature, and their fingerprint is related to the A and B sites of Fig. \ref{fig1:RT_sb2}. Notably, the main features associated with nearest- and next-nearest-neighbor Sb2 nuclei, as obtained by our calculations for low Sn doping, remain valid across the entire doping range investigated here by NQR, $i.e.,$ up to $x=0.65$.

Extending a similar accurate computational analysis to the high-temperature phase of these materials requires more advanced methods to estimate the free energy of the system as a function of temperature, such as molecular dynamics or the self-consistent stochastic harmonic approximation \cite{Park2023, alkorta2025symmetrybrokenchargeorderedgroundstate}. These approaches are computationally demanding and challenging to implement on the supercell method used in this study.  We use a simpler approach, described in the supplementary material, that yields the same qualitative results discussed above for low-temperature case.

It should be noted that these satellite peaks in NQR spectra appear to be robust and ``survive'' the CDW transition, supporting the conclusion that these features originate from the local distortion caused by the steric effect of Sn substitution.  

\begin{figure*}
\includegraphics[width=0.96\textwidth]{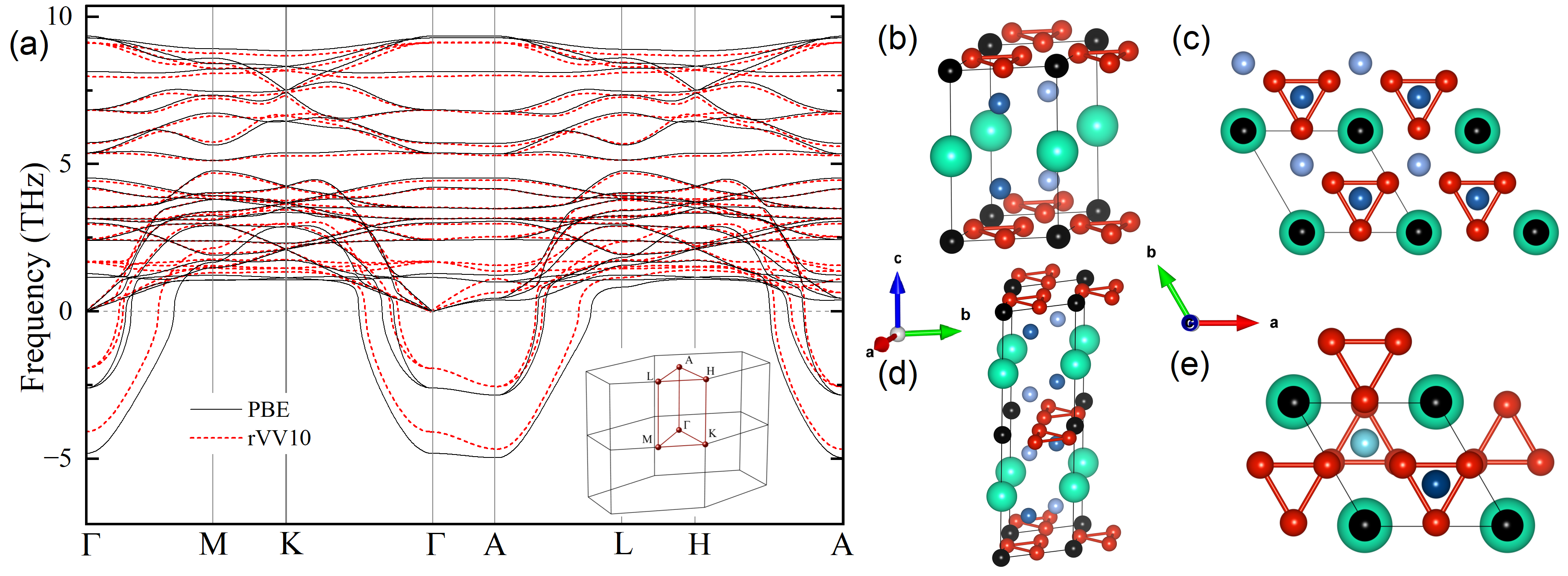}
\caption{(a) Calculated phonon dispersion curves for 100\% Sn substitution at the in-plane site of hexagonal CsV$_3$Sb$_4$Sn with PBE and rVV10 interactions. Phonon softening is observed at $\Gamma$ and $A$ high-symmetry points. The inset shows the Brillouin zone along the main symmetry directions. (b-c) show the newly identified $\Gamma$ crystal structure due to the softening of the zone-centre phonon mode in the $bc$ and $ab$ plane, respectively, creating the V-triangles and two symmetry-inequivalent Sb2 sites (light and dark blue spheres). (d-e) display the $A$ crystal structure, which has the same in-plane arrangement of V atoms, but staggered along the $c$-axis with a $\pi$-phase between the adjacent layers.}
\label{fig6:phonon_cvssn} 
\end{figure*}
Since the NQR technique is one of the most powerful tools to detect the character of the charge ordering of \avs kagome metal \cite{ilija_manuscript,PhysRevResearch.5.L012017,PhysRevB.107.184106,luo2022possible,Mu_2022}, a precise assignment of the spectral features is essential for an accurate interpretation of the data.
These results provide a strong basis for the interpretation of NQR measurements in doped kagome metals.  In the next section, we consider the fully Sn-substituted idealized system in order to examine the stability of the charge-ordered configuration.

%\subsubsection{Dynamical instability of CsV$_3$Sb$_4$Sn}
\subsection{Ground state of CsV$_3$Sb$_4$Sn}

To gain insight into the behavior at high doping levels, we also considered the end member of the series,  \cvss with $x=1$, where all in-plane Sb atoms are replaced by Sn, in a configuration-disorder-free picture.

At this concentration, Sn introduces a significant level of hole doping in the system \cite{CVSSn}, therefore affecting the Fermi-surface instabilities and possibly altering the electron-phonon coupling that stabilizes the TrH phase in the pristine compound. We therefore analyze the dynamical stability via phonon dispersion of the kagome structure for $x=1$ by performing $ab$-$initio$ calculations.
%The high-temperature hexagonal phase of \avs is characterised by instabilities at the high-symmetry $M$ and $L$ points \cite{dftcdw2,PhysRevMaterials.5.L111801, PhysRevB.104.214513}. By careful inspection of all symmetry allowed structures obtained from the softening phonon modes, the so-called staggered tri-hex structure is found to be the most stable one \cite{}. 
%To identify the equilibrium lattice structure of CsV$_3$Sb$_4$Sn, where Sn replaces the in-plane Sb atom, we proceed in a similar fashion.
Fig. \ref{fig6:phonon_cvssn} shows the vibrational spectrum for the hexagonal unit cell ($P6/mmm$), CsV$_3$Sb$_4$Sn \footnote{See SM for computational details.}. Imaginary (negative) phonon modes are identified at $\Gamma$ and $A$ high-symmetry points and, interestingly, the instabilities at the $M$ and $L$ points reported for the pristine \cvs are no longer present \cite{PhysRevMaterials.5.L111801, PhysRevB.104.214513,PhysRevLett.127.046401}. The phonon polarization vector corresponding to the imaginary mode at the $\Gamma$ point is connected to the stabilization of a three-V-atom triangle distortion depicted in Fig. \ref{fig6:phonon_cvssn}(b-c) in the  $bc$ and $ab$ planes, respectively, forming a V-trimerized phase. This structure is also characterized by two inequivalent out-of-plane Sb2 atoms, inside and outside the V-trimers (shown by light and dark blue spheres). 
%This structure is stable and reduces the total energy by 26 meV/u.c. with a GM4- mode stabilizing a low-symmetry 187 space group phase. 

In addition, the $A$-point instability produces a $\pi$-phase shift between adjacent layers along the $c$-axis of the structure, eventually resulting in an energy reduction of 3 meV/atom with respect to the hexagonal lattice. This lowest-energy structure is shown in Fig.~\ref{fig6:phonon_cvssn}(d-e) along the $bc$ and $ab$ planes, highlighting opposite shifts of V atoms in adjacent layers, creating a $\pi$-phase V-trimerized structure.
Detailed structural information, including lattice parameters, atomic positions, and the distances between the vanadium atoms in the triangular motif, is summarised in \mbox{Table \ref{tab:structure_info}}.

This suggests that the fully substituted $x = 1$ CsV$_3$Sb$_4$Sn system exhibits a lattice instability characterized by V-trimers, which lies extremely close in energy to its $\pi$-shifted phase along the $c$-axis. These instabilities are completely different from the TrH- and SoD-type reconstructions reported for the pristine kagome systems, highlighting a modified energy landscape of competing structural phases in the doped \mbox{kagome lattice}. 

\begin{table}
    \centering
    \begin{tabular}{c| c| c}
          & $\Gamma$ & $A$\\ \hline
      SG &  $P-6m2$ (187) & $Cmcm$ (63) \\
       $a$  &5.603  & 5.603 \\
       $c$  & 9.188 &  18.354 \\ 

& & \\
$\Delta E$ & 26 & 27 \\
   & & \\      
       d$_{V,tri}$ & 2.621 & 2.620 \\
       d$_{V,atri}$ & 2.981  & 2.982 \\
   & & \\    
     Cs   & 0.0,  0.0,     0.5000   ($1d$) &  0.0,  0.0,  0.2500  ($2a$) \\
   V  &    0.5106,   0.4893, 0.0   ($3j$) & 0.5050,  0.4972,  0.0   ($2c$)\\
   V & -- & 0.5050,  0.0295,  0.0   ($4g$)\\ 
    Sb  &   0.6666,   0.3333,  0.2512  ($2i$) & 0.6609,  0.3414,  0.1258  ($4f$)\\
    Sb  &   0.3333,  0.6666,  0.7609 ($2h$) &  0.6602,  0.3422,  0.6199  ($4f$) \\
    Sn  &   0.0,  0.0,  0.0  ($1c$) &  0.0,  0.0,  0.0   ($2c$)\\

    \end{tabular}
    \caption{Details of the equilibrium structures, labeled $\Gamma$ and $A$, obtained for CsV$_3$Sb$_4$Sn from estimated dynamical instabilities. $a$ and $c$ are the lattice parameters in \AA.  $\Delta E$ is the energy relative to the pristine hexagonal phase in meV per unit cell. The d$_{V,tri}$ is the distance between V atoms forming the triangle (in \AA), and d$_{V,atri}$ is the distance between V atoms outside the triangles, while in the pristine \cvs hexagonal lattice, the V-V distance is 2.762 \AA. }
    \label{tab:structure_info}
\end{table}

%The frequency obtained at the gamma point for the V-trimerized phase are all real and reported in the SM. Unfortunately, the computation of the phonon modes for the \mbox{$\pi$-phase} trimerized structure requires excessive computational resources.

%\textcolor{red}{How do I identify which symmetry this instability breaks? Also it should be breathing in mode, but I didn't find another structure with the breathing-out deformation. Is this deformation is only possible in Sn, so this is high-temperature, what about in CDW state? If we started from the CDW structure, we would still get this deformation. Thus it means there would be no CDW state for the doping? Should we also calculate the .band structure and Fermi surface structure for this new structure observed?}

\section{Discussion and Conclusion}

We investigated Sn-doped \cvs using a combined experimental and theoretical approach based on NQR measurements and first-principles calculations in order to study the microscopic structural configuration at both high and low temperature.
Sn doping yields additional satellite peaks in the NQR spectra, which are attributed to the local perturbation induced by a Sn impurity at the Sb1 site, propagating beyond nearest-neighbor shells, as evidenced by both the binomial trend of the experimental spectral weight (Fig.~\ref{fig2:reltaive weight}) and the DFT simulations at low doping. Surprisingly, the same set of additional satellite features persists below the CDW transition (Fig.~\ref{fig:impurity}), indicating an impurity-induced perturbation that is qualitatively reproduced by our computational analysis.  

At intermediate Sn concentrations, the NQR spectra retain similarities to the low-doping case, reflecting the effect of local distortions, while clearly deviating from a simple binomial description, revealing the presence of competing structural configurations in the doped kagome lattice.  In this regime, a more complex scenario is at play, where compositional disorder due to inequivalent arrangements of Sn on the Sb sublattice and multiple closely spaced electronic states are present. 
Recently, intermediate and low doping results have been analyzed in the framework of quenched disorder introduced by Sn impurities, which can pin and stabilize the CDW fluctuations, resulting in short-range static CDW domains that persist up to room temperature, even beyond the doping range where long-range CDW exists~\cite{deng2026observation}. This has been observed using ultrafast time‑resolved reflectivity~\cite{deng2026observation} and in recent NQR work~\cite{ilija_manuscript}. 

%Below the CDW transition, additional competing interactions are expected to emerge.  The breaking of  electronic symmetry in the doped kagome system may results from these interactions and local perturbations introduced by doping, as discussed in detail in the recent manuscript \cite{ilija_manuscript}.

To gain insight into the high‑doping limit, we examined the fully substituted compound CsV$_3$Sb$_4$Sn, as an idealised computational case. Our phonon dispersion calculations reveal a distinct kagome lattice instability characterized by the formation of vanadium trimers staggered along the $c$-axis, with an energy gain larger than that observed for the TrH structure of the pristine compound. These findings indicate that, in this theoretically fully substituted limit, Sn substitution induces a modified energy landscape with competing low-energy structural phases.

%In conclusion, we show that chemical pressure plays a key role in addition to the fermi level tuning and needs to be take into consideration.
To conclude, our results indicate that, in addition to Fermi-level tuning, the Sn substitution drives a complex interplay of local lattice distortions and competing structural instabilities in the \cvs system, providing a microscopic basis for understanding the rich and unconventional phase diagrams observed in doped kagome systems.

%\textcolor{red}{is there any observation of the pinning by disoder or lattice defcets creating the Sod or TrH deformation in the lattice? or any external factor that stabilises the deformation.How do we understand the fact this is not due to any kind of external strains, as these systems are highly sensitive to the strain.}

\section*{Acknowledgments}
We thank Andrew Cupo for the fruitful discussion.
AK and PB acknowledge the computational support by ISCRA initiative of CINECA with project IsCb6\_TRSBKS and CNHPC\_1570115. AK acknowledges funding support by PRIN project 202243JHMW.
Work in Parma was funded by the PNRR MUR Project No. ECS-00000033-ECOSISTER.
AC acknowledges support from PNRR MUR project PE0000023-NQSTI. DDS acknowledges MUR funding within the FIS2 (n. 1236, 01-08-2023) Project no. FIS-2023-00144 (CUP J53C25001880001). AC and DDS acknowledge the Gauss Centre for Supercomputing e.V. (\url{https://www.gauss-centre.eu}) for funding this project by providing computing time on the GCS Supercomputer SuperMUC-NG at Leibniz Supercomputing Centre (\url{https://www.lrz.de}). SDW and ADS acknowledge support via the UC Santa Barbara NSF Quantum Foundry funded via the Q-AMASE-i program under award DMR-1906325. VFM and IKN gratefully acknowledge support from the National Science Foundation via grant No. DMR-1905532 and funds from Brown University and University of Bologna.

\bibliography{ref}

\end{document}

\setcounter{section}{0}
\renewcommand{\thesection}{S\arabic{section}}
\setcounter{secnumdepth}{3}
\setcounter{figure}{0}
\renewcommand{\thefigure}{S\arabic{figure}}

\setcounter{table}{0}
\renewcommand{\thetable}{S\arabic{table}}

\def\rvs {RbV$_3$Sb$_5$\xspace}
\def\cvs {CsV$_3$Sb$_5$\xspace}
\def\kvs {KV$_3$Sb$_5$\xspace}
\def\cvss {CsV$_3$Sb$_{5-x}$Sn$_x$\xspace}
\def\V {$^{51}$V\xspace}
\def\Sb {$^{121}$Sb\xspace}
\textit{}
%Order-by-disorder: 
\title {Competing lattice structures induced by Sn substitution in \cvs - Supplementary Material}
\author{\hspace{1mm}Anshu Kataria}
\email[]{kataria96anshu@gmail.com}
\affiliation{Dipartimento di Scienze Matematiche, Fisiche e Informatiche, Universit\`a di Parma, I-43124 Parma, Italy}

\author{\hspace{1mm}Ilija K.~Nikolov}
\affiliation{Department of Physics, Brown University, Providence, Rhode Island 02912, USA}

\author{\hspace{1mm}Armando Consiglio}
 \affiliation{Dipartimento di Fisica e Astronomia ``A. Righi'', Universit\`a di Bologna, I-40127 Bologna, Italy }
\affiliation{Istituto Officina dei Materiali, Consiglio Nazionale delle Ricerche, Trieste I-34149, Italy}

\author{\hspace{1mm}Giuseppe Allodi}
\affiliation{Dipartimento di Scienze Matematiche, Fisiche e Informatiche, Universit\`a di Parma, I-43124 Parma, Italy}

\author{\hspace{1mm}Ginevra Corsale}
\affiliation{Department of Physics, Brown University, Providence, Rhode Island 02912, USA}
\affiliation{Dipartimento di Fisica e Astronomia ``A. Righi'', Universit\`a di Bologna, I-40127 Bologna, Italy }

\author{\hspace{1mm}Andrea Capa Salinas} 
\affiliation{
Materials Department, University of California Santa Barbara,
Santa Barbara, California 93106, USA}

\author{\hspace{1mm}Stephen D. Wilson}
\affiliation{
Materials Department, University of California Santa Barbara,
Santa Barbara, California 93106, USA}

\author{\hspace{1mm}Domenico Di Sante}
 \affiliation{Dipartimento di Fisica e Astronomia ``A. Righi'', Universit\`a di Bologna, I-40127 Bologna, Italy }

\author{\hspace{1mm}Vesna F. Mitrovi{\'c}}
\affiliation{Department of Physics, Brown University, Providence, Rhode Island 02912, USA}
\affiliation{Brown Center for Theoretical Physics and Innovation, BCTPI, Brown University, Providence, RI 02912-1843, USA}

\author{\hspace{1mm}Samuele Sanna}
 \affiliation{Dipartimento di Fisica e Astronomia ``A. Righi'', Universit\`a di Bologna, I-40127 Bologna, Italy }

\author{\hspace{1mm}Pietro Bonf\`a}
\email[]{pietro.bonfa@unimore.it}
\affiliation{Dipartimento di Fisica, Informatica e Matematica, Universit\`a di Modena e Reggio Emilia, Via Campi 213/a, 41125 Modena, Italy}
\affiliation{CNR-NANO S3—Istituto Nanoscienze, I-41125 Modena, Italy }

\date{\today}

\maketitle
\section{Experimental details}

Zero-field low- and room-temperature $^{121}$Sb nuclear quadrupolar resonance (NQR) experiments were performed on a high-purity  \cvss powder at the University of Parma, using a state-of-the-art home-built phase-coherent spectrometer. The powder sample was mounted on the resonant circuit probe. A standard spin echo sequence was used to obtain the frequency spectrum, with high signal averaging providing a high signal-to-noise ratio.  The spectra were recorded by sweeping the frequency and combining the individual spectra obtained by the echo sequence. 
%\textcolor{red}{should I mentioned here directly that we took the low-T from the ref. or because I don't know where the low-T data is acquired in parma or somewhere else}.

The NQR frequency of a nucleus with spin $I$ $>$ 1/2  and quadrupole moment $Q$ is $\nu_{Q} =  \frac{3e^2qQ}{2I(2I-1)h}$, where $eq = V_{zz}$ is the largest component of the EFG tensor in the principal axis system with $|V_{xx}| \leq |V_{yy}| \leq |V_{zz}|$, and $\eta=(V_{xx} - V_{yy})/ V_{zz}$ is the asymmetry parameter of the EFG.

Of interest for the present work is the $^{121}$Sb nuclei, with spin $I$ = 5/2, which can exhibit two NQR transitions 5/2 $\leftrightarrow $ 3/2 and 3/2 $\leftrightarrow $  1/2. In this study, we focus only on the former transition.
When EFG tensor is axially symmetric ($\eta=0$), the two transitions have frequencies ${\nu }_{5/2 \leftrightarrow 3/2}=2{\nu }_{3/2\leftrightarrow 1/2}=2\nu_Q$. However, this is not always true for Sb NQR in these compounds: simulations below show that structural distortions raise $\eta$ to 0.2 near distorted regions, whereas far from Sn atoms $\eta$ remains $\sim 0.1$ or smaller.

\section{Computation details}

We perform first-principles density functional theory (DFT) simulations using the plane-wave (PW) based code QuantumESPRESSO. To approximate the exchange-correlation contribution, we opt for the Perdew-Burke-Ernzerhof (PBE) functional \cite{Perdew1996} and include van der Waals interactions through the rvv10 method \cite{rvv10}.
The electric field gradient (EFG) values are computed through the GIPAW code \cite{gipaw}, which requires tight convergence of the computational parameters. A cutoff of 70 Ry for plane waves and 600 Ry for charge density, together with a $14 \times 14 \times 8$ Monkhorst-Pack \cite{Monkhorst1976} grid in reciprocal space, is sufficient to achieve convergence.  This k-mesh is properly scaled in supercell calculations used to study the doped systems. For the pristine system \cvs, the reported crystal structure with space group P6/mmm \cite{kagome.first} and for the Tri-hexagonal structure, the $Fmmm$ \cite{PhysRevMaterials.6.015001}  are used to estimate the EFG. For the current study, we focus on the EFG values of the Sb2 atom; thus, the simulation parameters are chosen to ensure an accuracy of 0.1 MHz in the Sb2 resonance frequency, while converging the Sb1 frequency to this accuracy requires much larger calculation parameters. The PW pseudopotentials of the PSLibrary \cite{DalCorso_2014} are adopted for all simulations.

The phonon spectra for $x = 1$ system are obtained with the Phonopy code, considering only PBE and rvv10 interactions \cite{phonopy}. A $3 \times 3 \times 2$ supercell is used for this. The structure with 100\% Sn doping is initially fully relaxed, while constraining the hexagonal symmetry. The force convergence threshold is set to 10$^{-5}$ Ry/Bohr, and the GBRV ultrasoft pseudopotentials \cite{gbrv} are employed with a plane wave and charge cutoff of 50 Ry and 550 Ry, respectively. The reciprocal space is sampled with $15 \times 15 \times 10$ and is reduced accordingly in supercell calculations.

The frequencies of the  $\Gamma$ structure consisting V-trimerized are calculated using density functional perturbation theory with a $10 \times 10 \times 6$ k-mesh grid and a small force convergence of  10$^{-5}$ Ry/Bohr. 
The results are shown in Table ~\ref{gamma_freq}.

\section{Experimental Sb NQR Frequencies as a Function of Doping }

\begin{figure}
\includegraphics[width=0.50\textwidth]{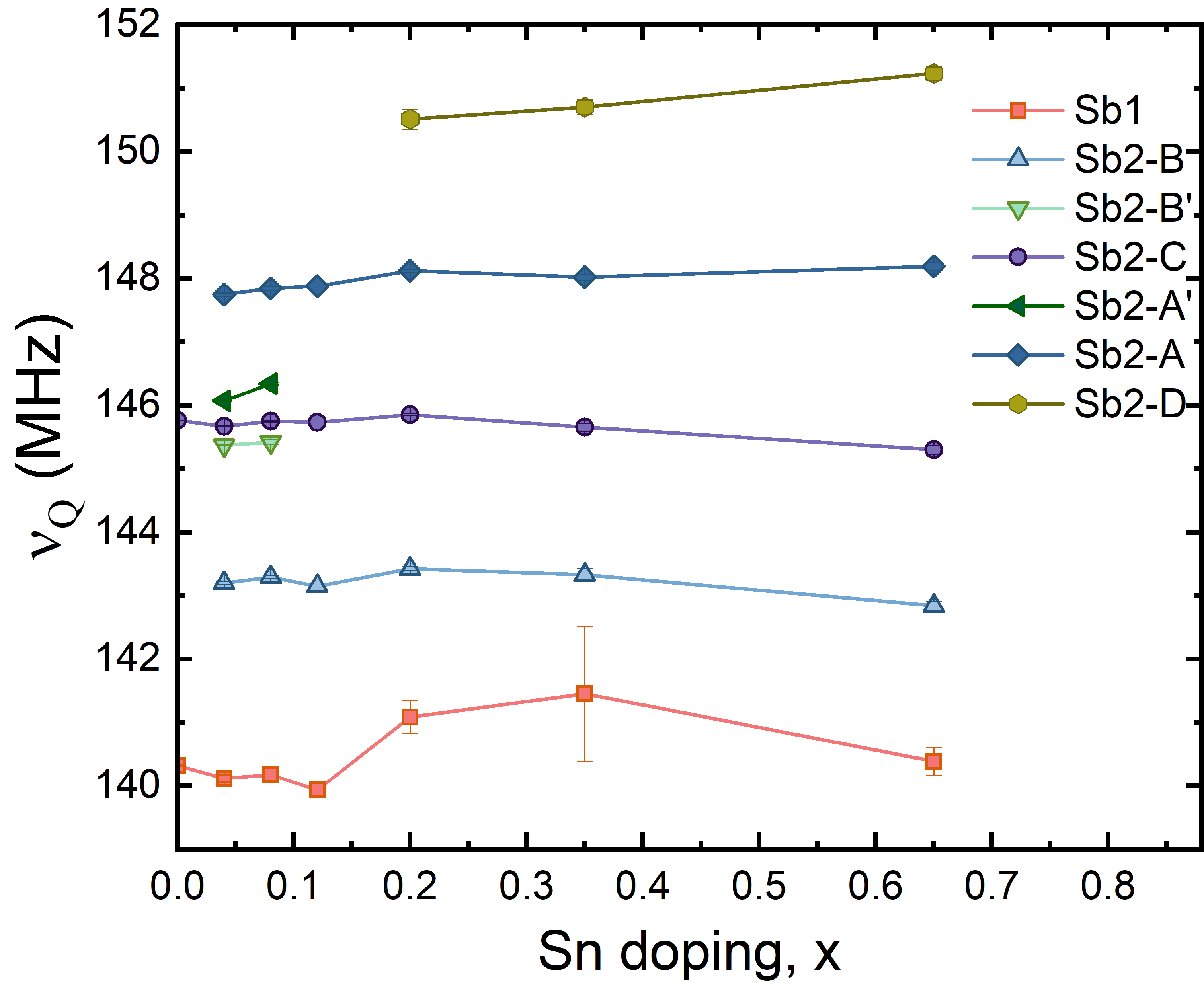}
\caption{Observed NQR frequencies variation with Sn doping, $x$ at room temperature. The color code and label of the NQR frequencies follow the same convention as mentioned in Fig. 1 of the main text. }
\label{smfig0:sm_0}
\end{figure}

The experimental NQR spectra for the entire doping series are fitted with a multiple Lorentzian function, as discussed in the main text. The center of each Lorentzian distribution is considered as the NQR frequency of the corresponding nucleus. The variation of the experimental NQR frequencies as a function of Sn doping $x$ is shown in Fig. \ref{smfig0:sm_0}. The color code and label follow the description of Fig. 1 in the main text. The in-plane Sb1 frequency (orange) exhibits minimal change upon doping, whereas the out-of-plane Sb2 NQR peak splits into multiple satellite peaks whose frequencies also remain largely independent of the increasing doping concentration up to the maximum value of $x=0.65$. 

\section{NQR frequencies for Sn-doping using Density Functional Theory}

\begin{table}
    \centering
    \begin{tabular}{|c|c|c|}
        \hline
     Holes &  $\nu_{Q,\text{Sb1}}$ (MHz)& $\nu_{Q,\text{Sb2}}$ (MHz)\\
       \hline
       0.00 &  115.1& 122.9\\
    
 0.10 & 118.3& 123.4\\
       
       0.20 & 121.6&123.9\\
       
       0.30& 125.0& 124.4\\

       0.40& 128.5& 124.8\\
       
       0.50 &132.2& 125.3\\
       \hline
       
    \end{tabular}
    \caption{Variation of $\nu_{Q}$ (in MHz) as a function of hole doping for a perfect crystal. }
    \label{tab:charge}
\end{table}

In order to understand the Sn occupancy site and its effect on the $^{121}$Sb NQR spectra, density functional theory calculations with two different approaches have been adopted. In the first approach, we remove the electrons from the unit cell along with the compensating uniform positive background in the perfectly periodic lattice to mimic the presence of a hole in the system. Sn doping acts as hole doping in the system. The change in resonance frequencies of Sb nuclei due to the presence of a hole is summarized in Table  \ref{tab:charge}. The in-plane Sb1 quadrupole frequency $\nu_Q$ varies by almost seven times more than the one for the out-of-plane Sb2 nuclei as the hole concentration increases, suggesting that electron removal has a stronger effect on the kagome plane, and leaves the charge distribution around the Sb2 nuclei only slightly affected.
This qualitative approach, however, cannot provide a microscopic description of the charge distribution around the Sb nuclei, since it preserves the original symmetry and neglects the lattice distortion introduced by Sn, and is therefore incapable of explaining the satellite peaks observed in experimental NQR spectra.

A second, more quantitative approach involves the substitution of Sn with Sb in supercells. In \cvs, Sn has two inequivalent sites in the hexagonal phase: in-plane and out-of-plane Sb atoms.  We consider Sn substitution at a single inequivalent site and observe a lower energy for the Sn at the Sb1-site configuration, consistent with previous observation \cite{CVSSn}. More refined relaxed structure calculations are discussed in the next section.

\begin{comment}
keeping all atoms fixed in the equilibrium positions of the pristine hexagonal lattice. Fig. \ref{smfig1:sm_1} shows the resonance frequency variation of Sb1 and Sb2 nuclei obtained in a 3$\times$3$\times$1 supercell as a function of the distance from the Sn atom for the two possible cases. 
In both cases it is noted that the perturbation induced by Sn is limited to the nearest neighbors.

For the case of Sn at the Sb2 site (Fig. \ref{smfig1:sm_1}(b)), there is a significant decrease in resonance frequency of Sb2 nuclei ($\nu_{Q,Sb2}$) within a distance of 4 \AA~from the Sn, while for Sn at Sb1 site (Fig. \ref{smfig1:sm_1}(a)) the effect is very limited.
This is a consequence of the fixed lattice geometry and of the localized nature of the charge redistribution, affecting only the kagome plane in Fig. \ref{smfig1:sm_1}(a) (in qualitative agreement with our previous result) and mostly localized on the impurity for the case in Fig. \ref{smfig1:sm_1}(b).
Both results are not compatible with our experimental NQR specta.
\end{comment}
%This is not compatible with the experimental  observation of almost symmetrical satellites around the Sb2 peak, though the unaffected central Sb2 peak (C-peak in Fig. 1 of the main text) can be understood from the Sn presence at the in-plane site. Further, the Sn substitution at the Sb1 site is found to be more energetically favorable compare to the other case in agreement with the previous report \cite{CVSSn}.
\subsection{Unconstrained Structural Relaxation }
\begin{figure*}
\includegraphics[width=0.70\textwidth]{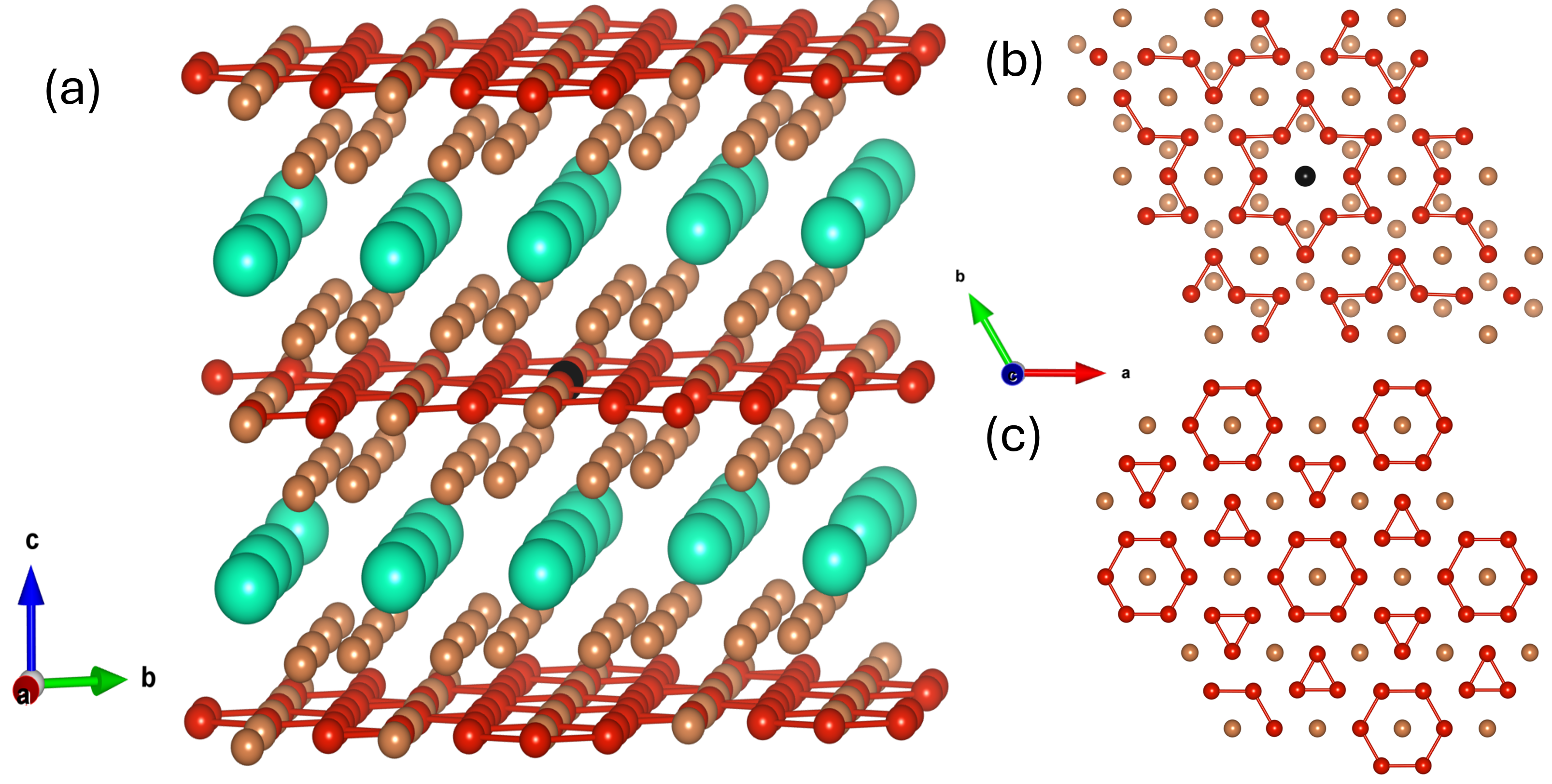}
\caption{ (a) Relaxed crystal structure of Sn-doped at in-plane Sb1 site for hexagonal \cvss. In the $ab$ plane, (b) a localised SoD pattern emerges around the Sn atom (black solid sphere) while the adjacent layer (c) exhibits a TrH structural motif. Cs, V and Sb atoms are depicted as cyan, red and brown spheres, respectively. }
\label{fig3:full_lattice}
\end{figure*}
\begin{figure*}
\includegraphics[width=0.60\textwidth]{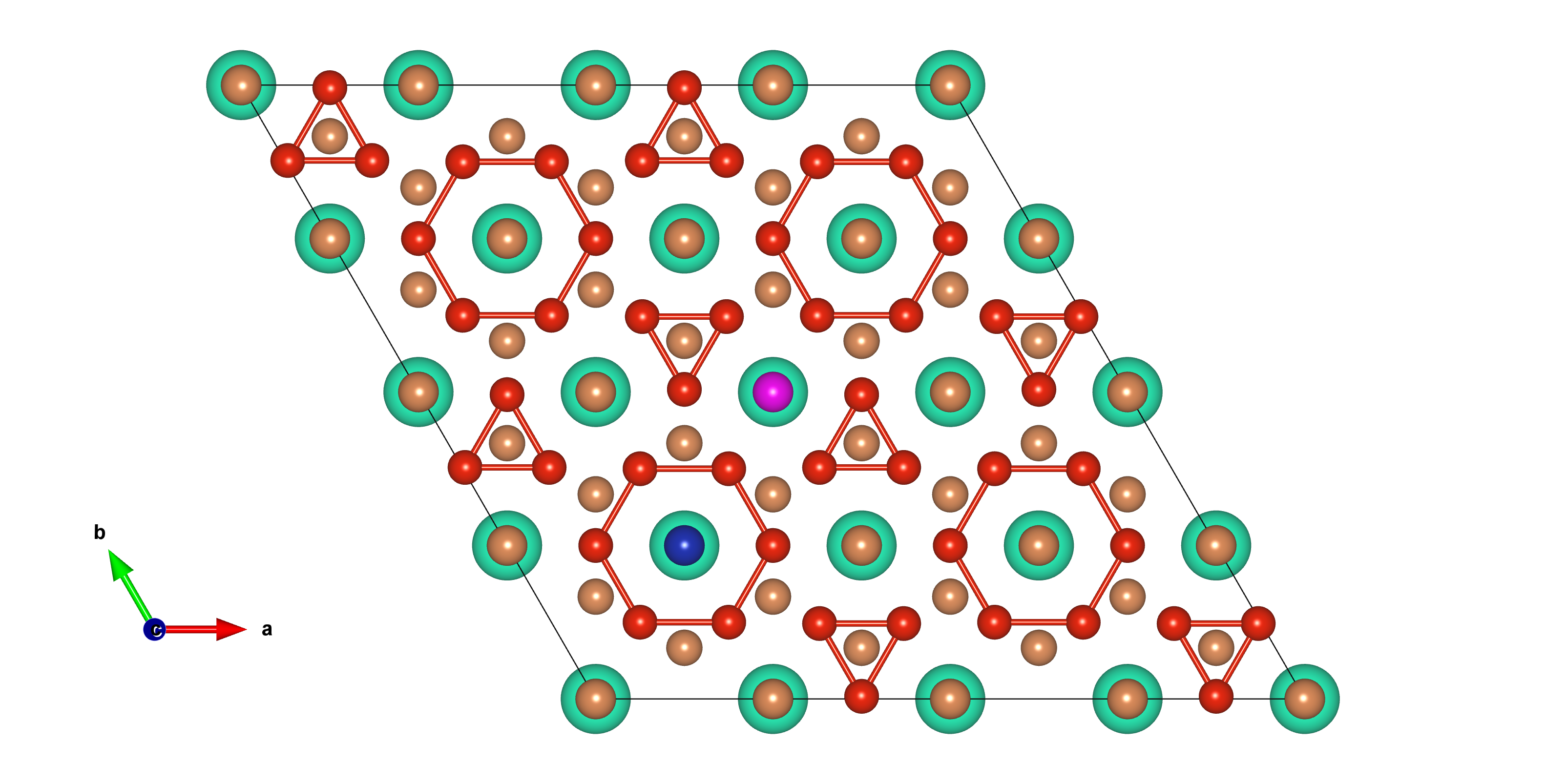}
\caption{The two possible Sn occupancies at the Sb1 site in the low-temperature TrH phase are represented by blue and pink solid spheres, corresponding to Sn inside and outside the V-hexagon, respectively. Other atoms representation is the same as in the last figure. }
\label{fig4:trh}
\end{figure*}
To refine our analysis, we performed unconstrained structural relaxation calculations of a hexagonal supercell, keeping the lattice parameters fixed while allowing only the atomic positions to relax. 
For the low-doping case $x \approx$ 0.021, we considered a $2 \times 2 \times 2$ supercell of hexagonal unit-cell with a single Sn atom located inside the V-hexagon, which forms part of the kagome plane. This approach provides a most controlled way to understand how a single dopant perturbs the RT structure.

The relaxed lattice is shown in Fig.~\ref{fig3:full_lattice}, where only the neighboring planes around the Sn impurity, represented by a solid black sphere, are displayed for clarity. As expected, the planes far from the impurity develop the so-called TrH structure  (Fig. \ref{fig3:full_lattice}(a) and (c)), which has been reported as the low-energy phase \cite{PhysRevLett.127.046401}. Nonetheless, close to the Sn atom, a SoD pattern is nucleated by the impurity  (Fig. \ref{fig3:full_lattice}(b)). The origin of this unexpected result can be understood as follows: the steric effect of Sn deforms the V-hexagon and drives the system towards the SoD pattern in the plane where the impurity resides, despite its higher energy. Taken together, these supercell simulations indicate that Sn substitution can lead to an exchange in the total energy of the two distortion patterns (SoD and TrH) of the kagome plane obtained for the pristine material. Consequently, the distortion effect due to doping in low-temperature TrH phase is examined in the following section. 

\subsection{Analysis of the Low-Temperature Phase }

Notably, in the TrH phase, Sn can occupy two inequivalent in-plane Sb1 sites: inside or outside the V-hexagon, corresponding to the Wyckoff sites $3f$ and $1a$, respectively. They are represented by blue ($3f$) and pink ($1a$) spheres in Fig. \ref{fig4:trh}. For the first case, Sn occupying the $3f$ site (inside the V-hexagon) stabilizes a SoD plane for $x=0.125$ in a 2x2x2 supercell. On the contrary, when Sn is placed outside the V-hexagon $1a$ site, the TrH structure remains the ground state. The corresponding predicted NQR spectra for this configuration are discussed in the main text. 

Again, this behavior is a consequence of the delicate balance between the steric effect of Sn and the spurious periodicity of Sn resulting from the supercell approach. These preliminary results highlight that modelling random Sn substitution at the Sb1 site is extremely challenging. 
%Furthermore, to enable a direct comparison with RT NQR spectra, constrained calculations were performed on the RT hexagonal phase.
%This suggests that local perturbation can influence the low-energy charge order, whether it manifests as TrH or SoD. 
%A second important conclusion obtained from the previous results is that, as already discussed in the literature \cite{Park2023}, the kagome planes are only weakly coupled.
%For these reasons, in order to keep the computational complexity tractable, we focus only on the low doping part of the experimental results and neglect the order along the $c$-axis (which has also been shown to have limited relevance in predicting EFGs at Sb atoms in the sister compound \rvs~\cite{RV3Sb5}).

\subsection{Constrained Relaxation of the Room-Temperature Phase }

To predict the NQR spectra at room-temperature, we adopted a simplified approach to constrain the lattice in the hexagonal symmetry: we used a $3\times 3 \times 1$ hexagonal supercell that conflicts with the formation of the ground state CDW structure characterized by the propagation vector $(\frac{1}{2},\frac{1}{2},\frac{1}{2})$ \cite{PhysRevLett.127.046401,jiang2021unconventional}.
Nevertheless, the supercell is sufficiently large to partially capture the structural distortions induced by the impurity through structural relaxation calculations. 

\begin{figure}
\includegraphics[width=0.45\textwidth]{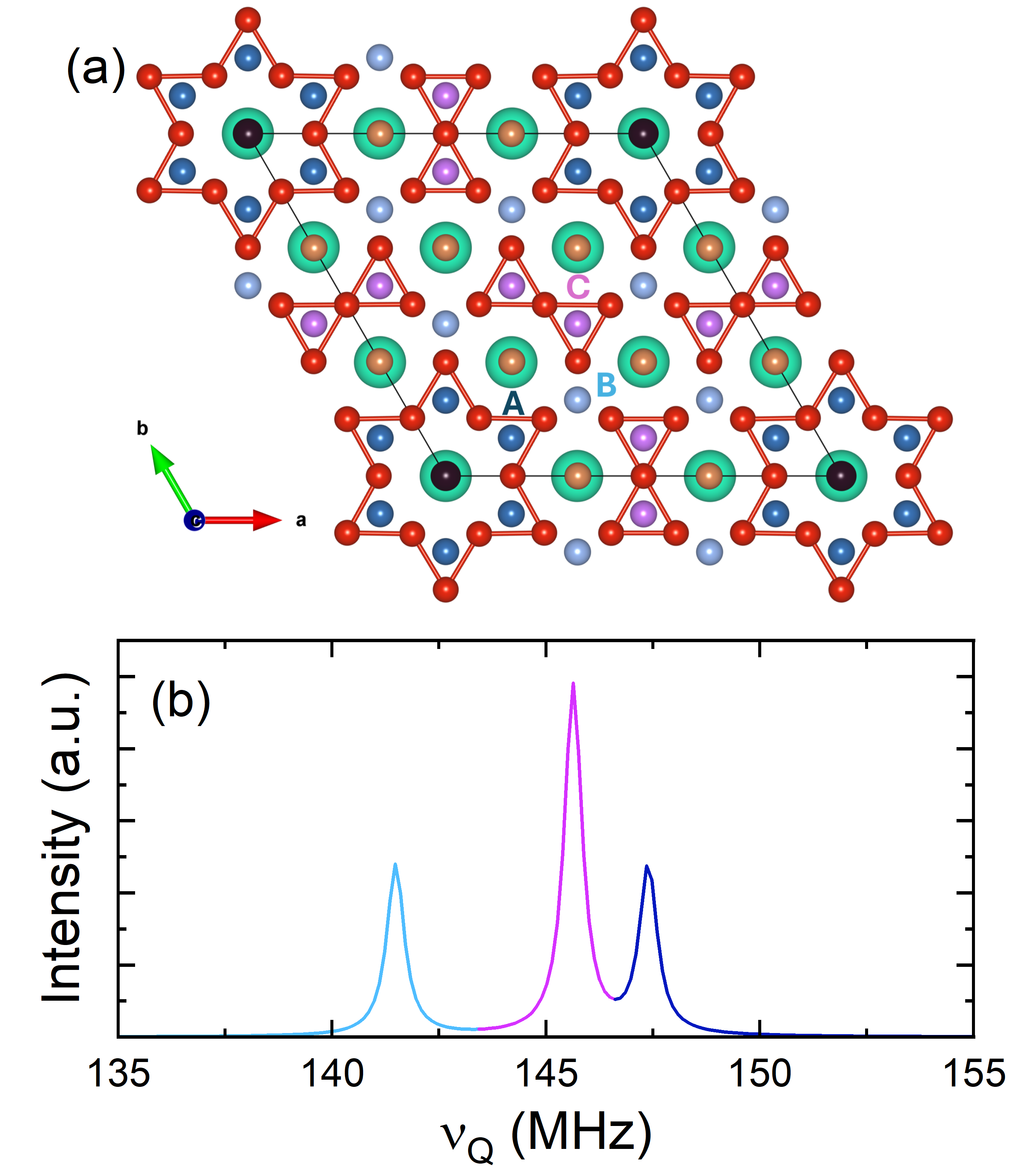}
\caption{In panel (a), the  3$\times$3$\times$1 relaxed supercell structure of the Sn-doped CsV$_3$Sb$_{5-x}$Sn$_x$ depicts the Star of David pattern with a black Sn atom at the center, Sb2 atoms with blue (dark and light) and purple spheres, in-plane Sb atoms with a brown sphere and a V atom with red spheres. The color of the two Sb2 sites, light and dark blue labelled as the A and B sites, is different from the purple Sb2 site part of the pristine cell. 
In panel (b), the predicted spectra for the above constrained lattice with sufficient Lorentzian distribution are presented. The colors and labels in the lattice are chosen to match the color assignment of Sb2-nuclei in panel (a). }
\label{fig4:sod}
\end{figure}
Following structural relaxation starting from the hexagonal lattice, we find that V atoms surrounding Sn are shifted from their equilibrium position, eventually leading to the formation of a Star of David: the NN V atoms (labeled V-NN) move outward from Sn in a radial direction, while the opposite displacement is observed for the next-NN (NNN) V atoms. Quantitatively, in our case, distances from Sn to the NN and NNN V atoms for the distorted (hexagonal) structures are $d_{\text{Sn, V-NN}}$ = 2.818 (2.762) \AA~ and $d_{\text{Sn,V-NNN}}$ = 4.691 (4.783) \AA, respectively. In a pristine kagome lattice acquiring a SoD distortion, the observed distances from the in-plane Sb atom at the hexagonal center are 2.772 (2.762) \AA~ and 4.764 (4.783) \AA ~~\cite{kagome.first}. 
%This is indeed the pattern leading to the formation of a Star of David.
As a consequence, two distinct local charge environments are created for Sb2 atoms: the Sb2 atoms closer to the Sn (12 sites shown as A in Fig.~\ref{fig4:sod}) are surrounded by V triangles that shrink, while the second set of Sb2 atoms (12 sites labeled B in Fig.~\ref{fig4:sod}) are closer to an expanding V triangle.
Finally, the sites labeled C are almost unaffected, but this is due to the limited size of the supercell.

%Further, the quadrupolar resonance frequencies for the doped, relaxed structure are determined.
Fig.~\ref{fig4:sod} (b) shows the DFT-predicted Sb2 NQR frequencies for 5/2 $\leftrightarrow $ 3/2 transition.
Also in this case, an appreciable shift of Sb2 NQR frequency by about $\pm$3 MHz is produced for the first and the second NN Sb2 from Sn atom, again suggesting that the NQR satellite peaks appearing with doping at RT are a consequence of the local deformation of the lattice induced by the Sn impurity.

%Overall, our results suggest that the satellite peaks appearing in the low-doping part of the NQR spectra originate from the local charge redistribution at the Sb2 sites caused by the distortion induced by Sn substitution, which surprisingly extends beyond the nearest neighbors, affecting out-of-plane Sb atoms as far as almost 7~\AA.

\section{Phonon analysis for CsV$_3$Sb$_4$Sn}

All calculated $\Gamma$-point phonon modes exhibit real frequencies for the V-trimerised structure, listed in Table \ref{gamma_freq}, indicating that the structure is dynamically stable. Notably, this structure is 10 meV/atom lower in energy than the SoD or TrH distortion in the $x= 1$ doped case.  

\begin{table}
    \centering 
    Frequency (MHz) \\
    \begin{tabular}{ c  c }\hline
    
        \hline
   -0.0434 &      3.4918 \\
    
   -0.0296 &  3.5536 \\
    
   -0.0150 &  4.1505 \\
   
    1.4843 &   4.5074 \\
   
    1.5036 &   5.7034 \\
   
    1.6531 &  5.7151 \\
    
    2.2985 &  7.4612 \\
   
    2.4311 &   7.6482 \\
    
    2.5628 &  7.6857 \\
    
    2.5369 &  8.1271 \\
    
    2.8805 & 8.7292 \\
    
    3.1135 & 9.0778 \\
    
    3.1664 & 9.0791 \\
    3.3436 & \\
          \hline
    \end{tabular}
    \caption{List of frequencies for the V-trimerised structure calculated at the $\Gamma$ point. No imaginary modes are observed except for a negligible contribution reported as negative frequencies for the three acoustic modes.}
    \label{gamma_freq}
\end{table}

\bibliography{ref}